\documentstyle[aps,prb,floats,epsf] {revtex}
\begin{document}

\advance\textheight by 0.2in

\draft

\twocolumn[\hsize\textwidth\columnwidth\hsize\csname@twocolumnfalse%
\endcsname

\title{Phase Diagram of the Two Dimensional Lattice Coulomb Gas}

\author{Pramod Gupta and S. Teitel }

\address{Department of Physics and Astronomy, University of Rochester, 
Rochester, New York 14627}

\date{\today}
\maketitle

\begin{abstract}
We use Monte Carlo simulations to map out the
phase diagram of the two dimensional Coulomb gas on a 
square lattice, as a function of density $\rho$ and temperature T. 
We find that the Kosterlitz-Thouless transition remains up to
higher charge densities than has been suggested by recent theoretical 
estimates.
\end{abstract}
]

The nature of phase transitions in the two dimensional 
neutral Coulomb gas (CG) has remained a topic of considerable interest.
The CG can be related via duality transformation to the 2D XY model,
and thus to superfluid and superconducting films.\cite{review}
The pioneering work of Kosterlitz and Thouless\cite{KT} (KT) showed that, at
low charge density, there is a second order transition from an 
insulating to a conducting phase, due to the unbinding of neutral 
charge pairs.  As the charge density is increased, several 
authors\cite{Minnhagen,Knops,Fisher,Friesen} 
have predicted that this KT transition should become first order.
Recently, Levin et al.\cite{Fisher}, using a modified Debye-H\"{u}ckel approach,
have estimated that, for a continuum
CG, the KT transition ends in a tricritical point at the surprisingly
low density of $\rho_{\rm c}\simeq 0.004/a^2$, where $a$ is the hard core
diameter of the charges.  

To investigate this issue, we report here on new Monte Carlo simulations
of the 2D CG on a square lattice, computing the
phase diagram as a function of density and temperature.  We make
sensitive tests of the nature of the transition, and conclude
that it remains second order up to densities much higher than
estimated by Levin et al.

Our work follows that of Lee and Teitel,\cite{Lee} with a few modifications.
We take for the Hamiltonian,
\begin{eqnarray}
  {\cal H}&=&{\scriptstyle {1\over 2}}\sum_{i,j}n_iG^\prime({\bf r}_i-{\bf 
  r}_j)n_j-u\sum_in_i^2 \nonumber\\ &+ &\sum_i(n_i^4-n_i^2)+
  V\left({2\pi P_x\over L}\right) +V\left({2\pi P_y\over L}\right) \enspace ,
\label{e1}
\end{eqnarray}
where the sums are over all sites $i$, $j$ of a 2D periodic $L\times L$
square lattice, with unit grid spacing.  
$G^\prime({\bf r})\equiv G({\bf r})-G(0)$, where $G({\bf r})$ 
is the solution to the lattice Laplacian with
periodic boundary conditions.  For $r\ll L$, $G^\prime({\bf r})\simeq -\ln 
r-{\scriptstyle {1\over 2}}\ln(8{\rm e}^{2\gamma})$, where $\gamma\simeq 
0.5772$ is Euler's constant.\cite{KT,JKKN}  $n_i=0,\pm 1,\pm 2\ldots$ 
are the integer
charges, and neutrality $\sum_in_i=0$ is imposed.  The third term 
in ${\cal H}$
tends to suppress charges with $|n_i|>1$, and is needed to
stabilize the system in the very dense limit.  Assuming therefore
that all charges satisfy $|n_i|\le 1$, so that $|n_i|=n_i^2$, the
second term is just $-u\rho L^2$ with $\rho=L^{-2}\sum_i|n_i|$ the charge
density.  Thus $u$ is the chemical potential.  The last two terms in ${\cal H}$
are effective boundary terms, which arise in the duality mapping
to the CG from the 2D XY model with {\it periodic} boundary 
conditions.\cite{Vallat,Young,Olsson}
Here ${\bf P}\equiv\sum_i n_i{\bf r}_i$ is the net dipole moment of the
charges, and $V(\phi)$ is the Villain function,\cite{Villain} 
$\rm{e}^{-V(\phi)/T}
\equiv \sum_{m=-\infty}^{\infty}\rm{e}^{-{1\over 4\pi T}(\phi-2\pi 
m)^2}$.  With these boundary terms, it is easy to 
compute\cite{Vallat,Olsson} the inverse
dielectric tensor $\epsilon^{-1}_{\mu\nu}$, which is the equivalent of the 
helicity modulus of the corresponding XY model,\cite{Ohta,Warren}
\begin{eqnarray}
  \epsilon^{-1}_{\mu\nu}&=& 2\pi\left\langle V^{\prime\prime}
  \left({2\pi P_\mu\over L}\right)\right\rangle\delta_{\mu\nu} 
  \nonumber\\&-&{2\pi\over T}\left\langle
  V^\prime\left({2\pi P_\mu\over L}\right)
  V^\prime\left({2\pi P_\nu\over L}\right)\right\rangle
  \enspace ,
\label{e2}
\end{eqnarray}
where $V^\prime$ and $V^{\prime\prime}$ are first and second
derivatives of $V$.  For an isotropic system, 
$\epsilon^{-1}_{\mu\nu}=\epsilon^{-1}\delta_{\mu\nu}$.
According to the KT instability criterion,\cite{KT} in the insulating phase
we must have the inequality $\epsilon^{-1}\ge 4T$.  This will
enable us to set an upper bound on the transition to the conducting
phase.

We carry out standard Metropolis Monte Carlo simulations for
various values of $u$ and $T$.
At each point an initial 20,000 MC passes are discarded to 
equilibrate with an additional 128,000
MC passes to compute averages. Errors are estimated from block
averages. 

\begin{figure}
\epsfysize=2.0truein
\quad\epsfbox{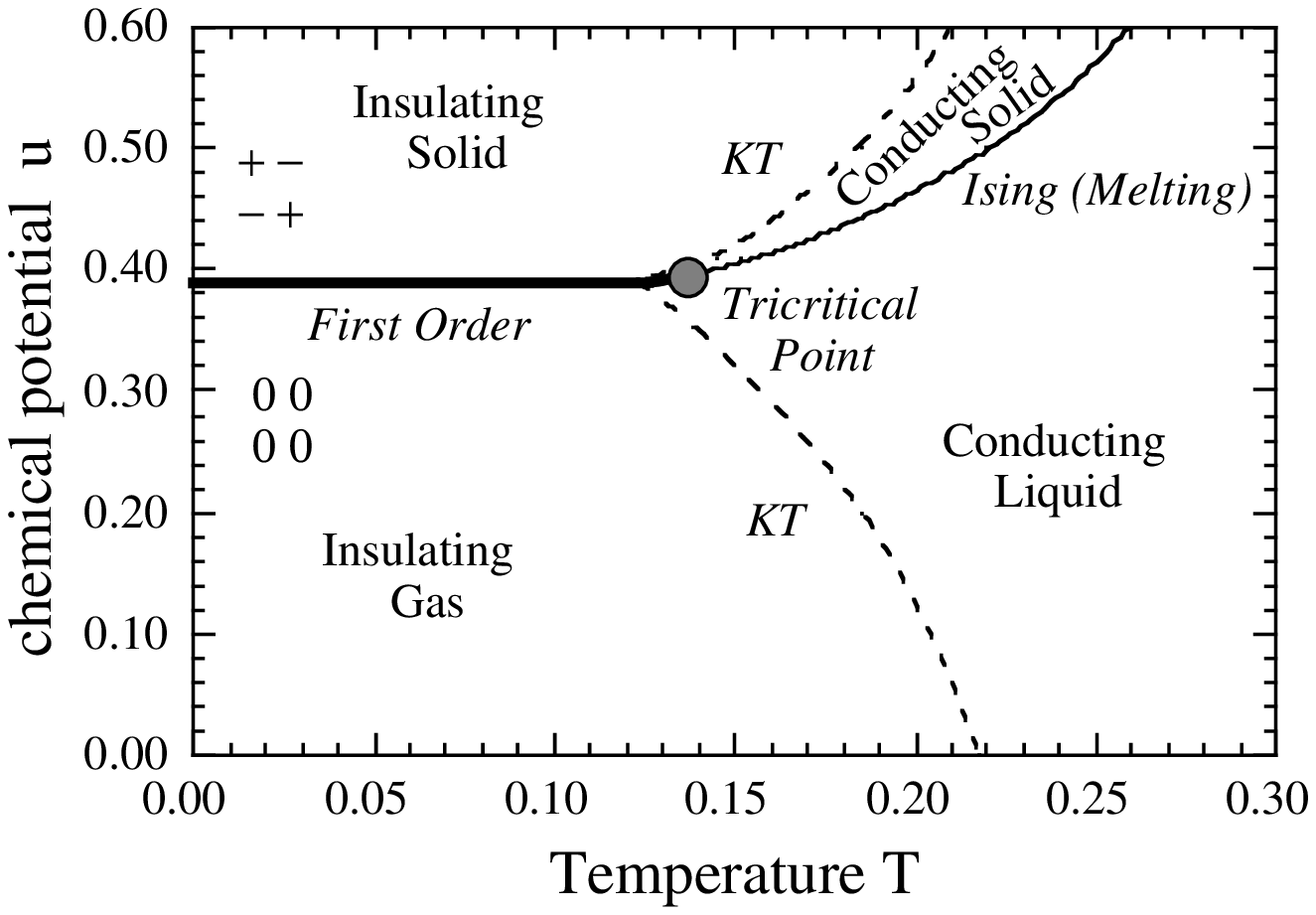}
\caption{Phase diagram of the CG in the $u-T$ plane.}
\label{f1}
\end{figure}

The $u-T$ phase diagram, as found previously by Lee and 
Teitel,\cite{Lee}
is shown in Fig.\,\ref{f1}. 
At low $u$, upon increasing $T$, there is a KT transition to 
a conducting liquid. At low $T$,
upon increasing $u$, there is a first
order transition to an insulating charge solid at $u_0=\pi/8$.
Increasing $T$ within this charge solid gives first a KT transition
to a conducting solid, followed by an Ising melting of the solid.
The two KT lines meet the first order line at the same $T^*\simeq 0.125$, 
slightly lower than
the tricritical point where the first order and Ising lines meet.

\begin{figure}
\quad\epsfysize=2.0truein
\epsfbox{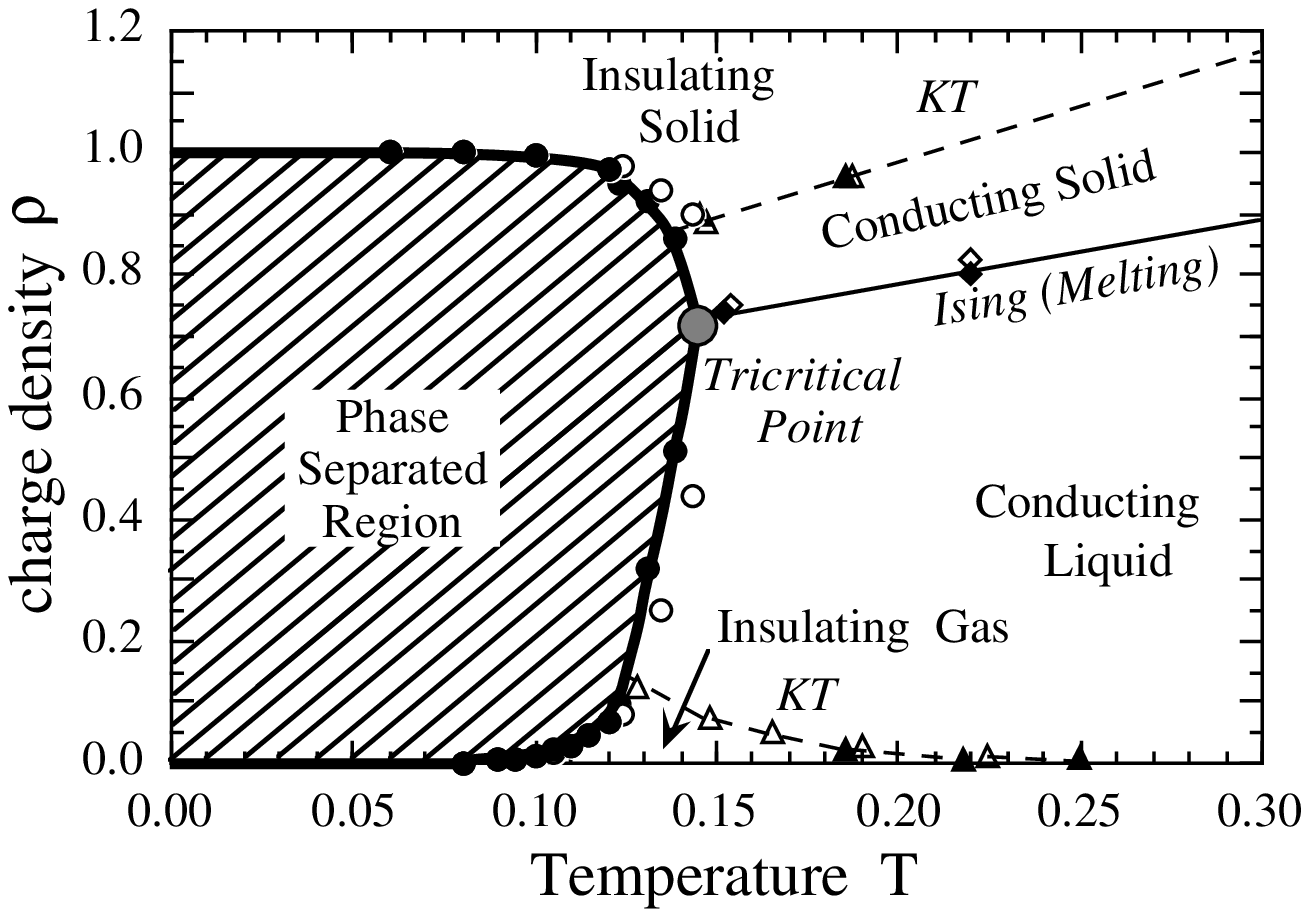}
\caption{Phase diagram of the lattice CG in the $\rho-T$ plane.  Open 
symbols are from simulations with system size $L=16$; solid symbols 
are from $L=32$.}
\label{f2}
\end{figure}

\begin{figure}
\epsfysize=1.8truein
\epsfbox{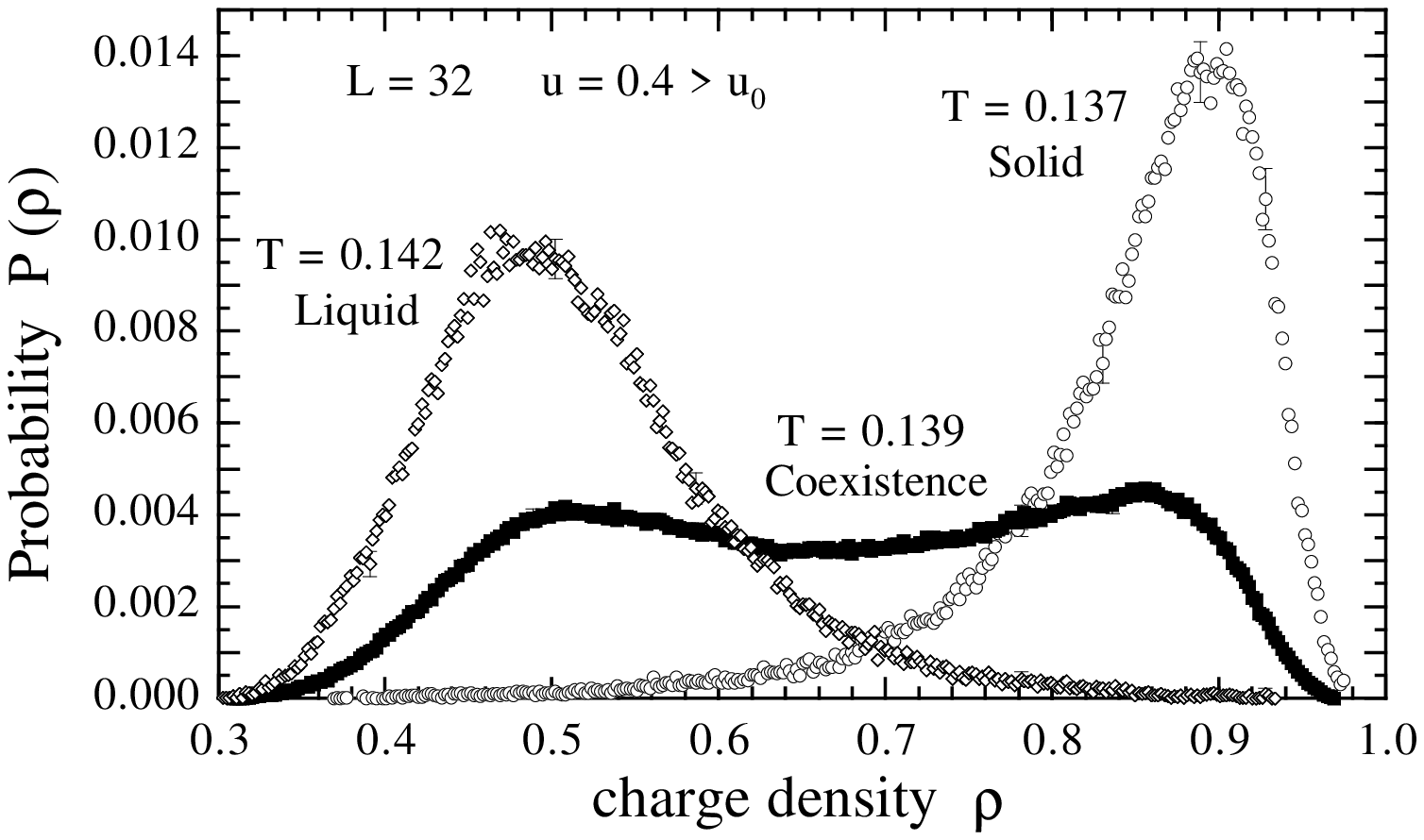}
\caption{Histograms of charge density $\rho$ for $u=0.4$ just above 
$u_0=\pi/8$.  At the coexistence boundary the histogram is bimodal}
\label{f3}
\end{figure}

In Fig.\,\ref{f2}, we present our new results for this phase
diagram in the $\rho-T$ plane.  
The coexistence boundary is determined 
as follows.  The low temperature branches are obtained by simulating 
with $u$ just above and just below $u_0$, measuring the average 
density $\rho$ as $T$ increases. 
To determine the boundary closer to the tricritical 
point, we simulate with fixed $u>u_0$, increasing $T$, and measuring
the histogram of the values of $\rho$ found at each value of $T$.
When one is in either the solid or the liquid phase, this histogram 
has a single peak.  However,
when one crosses the first order line, this histogram develops 
double peaks.\cite{LK}  The locations of the two peaks determine the 
densities of the two coexisting phases.  In Fig.\,\ref{f3} we show an 
example of such histograms.  Varying $u>u_0$ then maps out the rest of
the coexistence boundary.
In Fig.\,\ref{f2} we show the 
coexistence boundary found in this way from simulations with $L=16$ 
and $L=32$.  As is seen, and as is expected, our results near the tricritical 
point are limited by finite size effects.  However it is clear that 
the KT line at small $u$ joins the first order line at a density
$\rho\simeq 0.1$, much larger than the estimate of Levin et al.

\begin{figure}
\epsfysize=1.8truein
\epsfbox{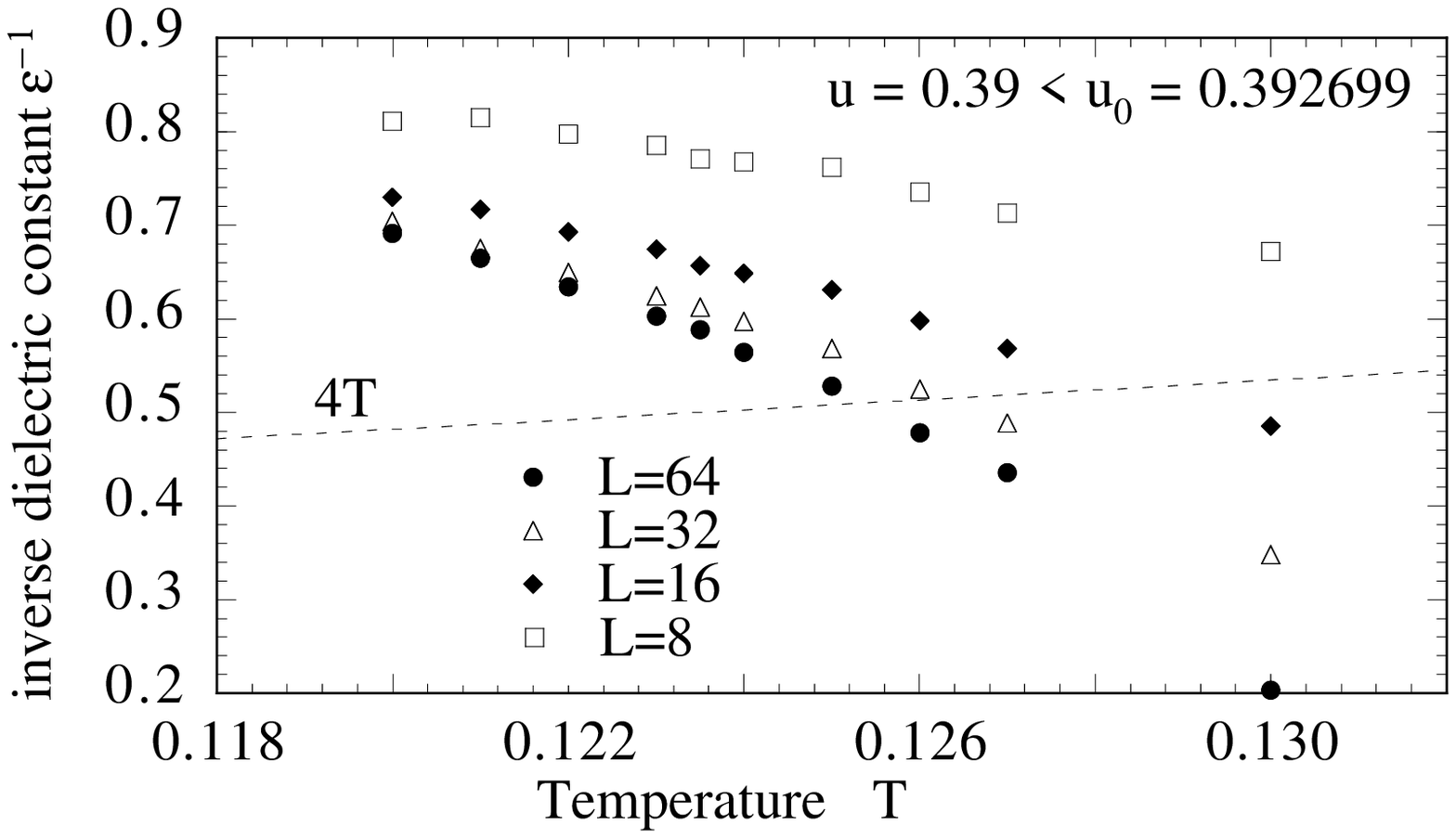}
\caption{Inverse dielectric function $\epsilon^{-1}(T)$ for various 
lattice sizes $L$, for $u=0.39$ just below $u_0=\pi/8$.  
The intersection with the line $4T$ gives the KT upper
bound on the transition temperature $T^*$.}
\label{f4}
\end{figure}

Next we verify that what we have called the ``KT'' line at small $u$ 
actually does remain a second order KT transition all the way up to the 
first order line separating the insulating gas and insulating solid.
To locate the transition temperature $T^*$ for $u=0.39$, just
below $u_0$, we compute $\epsilon^{-1}(T)$ for various $L$, using
Eq.\,(\ref{e2}).  Our results are shown in Fig.\,\ref{f4}.  The
intersection of these curves with the line $4T$ determines
the upper bound  $T^*\simeq 0.125$.  No hysteresis, or other suggestion 
of a first order transition was observed in $\epsilon^{-1}$. 

\begin{figure}
\epsfysize=3.6truein
\epsfbox{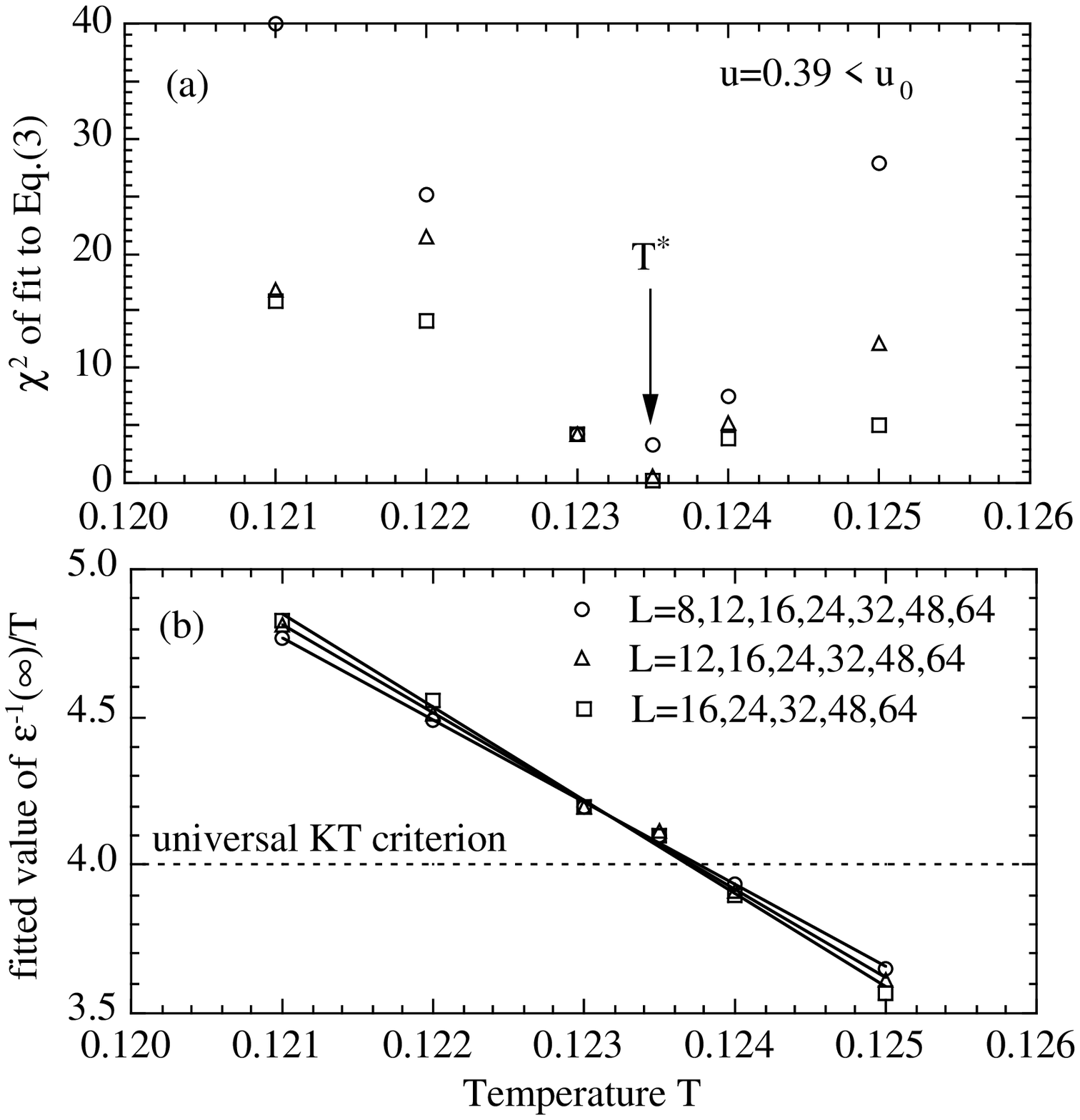}
\caption{Fitting of $\epsilon^{-1}(T,L)$ to Eq.(\ref{e3}). ($a$) shows 
the $\chi^2$ of the fit vs. $T$, ($b$) shows the value of the fitted
$\epsilon^{-1}(\infty)/T$ vs. $T$.}
\label{f5}
\end{figure}

As a more precise test that the transition is indeed KT--like, we use the 
finite size scaling procedure of Weber and Minnhagen \cite{Weber,note2}.  
Precisely at the KT transition temperature, the finite size dependence of 
$\epsilon^{-1}$ is given by,
\begin{equation}
   \epsilon^{-1}(L)=\epsilon^{-1}(\infty)\left[1+{1\over 
   2\ln L +c}\right]
\label{e3}
\end{equation}
Fitting data for various $L$ at fixed $T$ 
to Eq.(\ref{e3}), with $\epsilon^{-1}(\infty)$ 
and $c$ as free parameters, one determines $T_{\rm KT}$ as that temperature 
which gives the smallest $\chi^2$ error of the fit.  The fitted value 
of $\epsilon^{-1}(\infty)$ at the $T_{\rm KT}$ so determined, 
should then turn out to be precisely $\epsilon^{-1}(\infty)=4T_{\rm 
KT}$, so
as to obey the universal KT prediction.  We show the results of such
a fit below.  In Fig.\,\ref{f5}$a$ we show the $\chi^2$ of the fit 
versus temperature, where we have included sizes $L=8-64$, $12-64$, and
$16-64$ in the fit.  In each case, the minimum of $\chi^2$ occurs at
$T^*=0.1235$.  In Fig.\,\ref{f5}$b$ we show the corresponding fitted
value of $\epsilon^{-1}(\infty)/T$ versus temperature.  We see that
$\epsilon^{-1}(\infty)/T=4$ at essentially the same $T^*$ where $\chi^2$ has
its minimum.  This analysis strongly supports the transition as being
of the KT type.

\begin{figure}
\epsfysize=2.0truein
\epsfbox{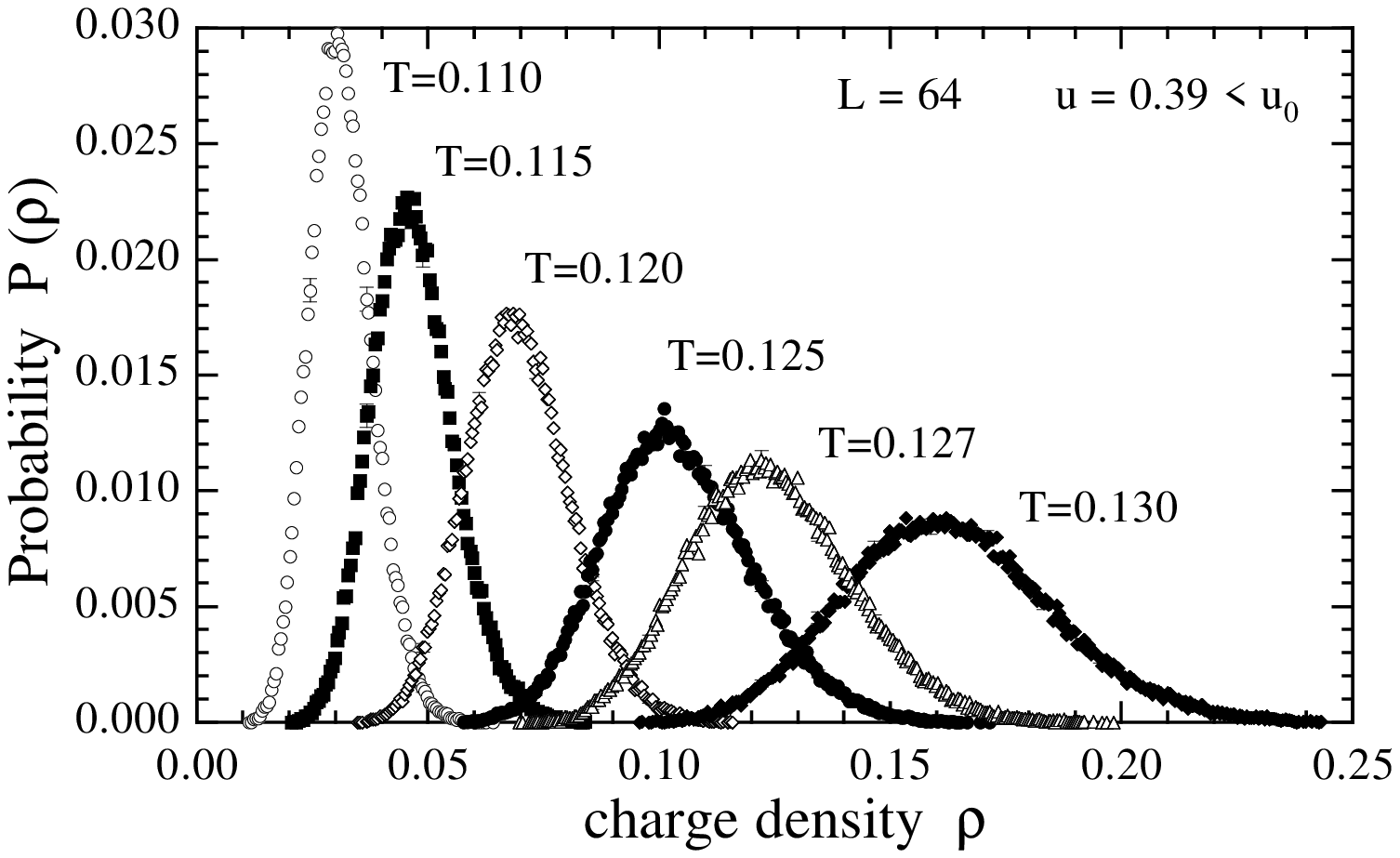}
\caption{Histograms of charge density $\rho$ for $u=0.39$ just below
$u_0=\pi/8$.  Histograms remain single peaked as one passes though the
transition at $T^*\simeq 0.1235$}
\label{f6}
\end{figure}

Finally, in Fig.\,\ref{f6} we 
show histograms of the density $\rho$ for $u=0.39$ and several values of $T$ 
passing through $T^*$, for the largest system size we have studied, 
$L=64$.  We have chosen the values of $T$ shown in  Fig.\,\ref{f6} so that the
histograms between neighboring values have a significant overlap.  As 
is clearly seen, all the distributions are single peaked.  There is no 
sign at all of the bimodal distribution that would characterize a 
first order transition.  While we cannot rule out the possibility of 
a weak first order transition with a $\xi(T^*)>64$, our results 
clearly suggest that the ``KT'' line at small $u$ does indeed remain a
second order Kosterlitz-Thouless transition all the way up to $u_0$.

Our lattice simulation has the advantage that it is easier to 
equilibrate at high densities than continuum simulations.  
However the presence of the discrete lattice does have
a strong influence on the location of the phase boundaries.  
In particular, the coexistence boundary of Fig.\,\ref{f2}
at small $T<0.1$ is well
described by a simple model of excited isolated dipoles and quadrupoles.
For the lower branch we get
$\rho_-(T)={1 \over L^2}(2N_{d} {\rm e}^{-\beta E_{d}} + 4N_{q}
{\rm e}^{-\beta E_{q}})$,
where $E_d$ and $E_q$ are the excitation energies of an isolated 
dipole and quadrupole respectively, and $N_d=4L^2$ and $N_q=2L^2$ 
are the number of 
ways of putting them down on the square lattice.
For the upper branch we get
$\rho_+(T)=1-{1 \over L^2}(2N^\prime_{d} {\rm e}^{-\beta 
E^\prime_{d}} + 4N^\prime_{q}
{\rm e}^{-\beta E^\prime_{q}})$ where $E^\prime_{d}$ and 
$E^\prime_{q}$ are the energies for removing an isolated dipole and 
quadrupole, and $N^\prime_d=2L^2$ and $N^\prime_q=L^2$ are the 
number of ways this may be done.  Clearly these expressions will change
with the geometry of the discretizing lattice, or if a continuum is 
used.  Nevertheless, at the very low densities $\rho\sim 0.004$ and 
high temperatures $T\simeq 0.25$ where Levin et al. estimate a 
tricritical point, we would be very surprised if the lattice is 
qualitatively different from the continuum.

Using a discrete lattice also has the effect that it
tends to stabilize the charge solid phase above $u_0$
to high temperatures.  Indeed, our charge solid only melts
after it has already become conducting via a KT transition arising 
from the excitation and diffusion of vacancies throughout the solid.
The present model does not possess any first order transition from
an insulating to a conducting phase.
In contrast, recent simulations\cite{Ork,Wallin} of the CG in a flat {\it 
continuum} with periodic boundary conditions find that the charge solid 
phase is melted at any finite temperature.  Here, a first order line 
separates the insulating gas from a dense conducting liquid, ending 
at a critical point at relatively low temperature and high density:
$(T_c,\rho_c)=(0.056,0.21)$ according to Ref.\,\onlinecite{Ork}, and
$T_c=0.032$ according to Ref.\,\onlinecite{Wallin}.  Similar results 
were found earlier for the continuum CG on the surface of a 
sphere:\cite{Caillol}
$(T_c,\rho_c)=(0.087,0.11)$.  In these models, the KT line ends either 
at, or near this critical point.

Although the geometry of the the CG system clearly affects the location
of the end of the KT transition line, it is interesting
to compare our results for the square lattice with the predictions of 
the continuum self-consistent screening theory of Minnhagen and 
Wallin.\cite{Minnhagen}  To do 
so, it is necessary to note\cite{KT,JKKN}
that if the interaction $G(\vec{r})$ on the lattice is chosen so as
to asymptotically match the continuum $-\ln r$ as $r\to\infty$, 
then the lattice CG with $u=0$ acts like a continuum model with a
chemical potential $\mu_0=-{1\over 4}\ln(8{\rm e}^{2\gamma})\simeq
0.8085$. Thus
a chemical potential $u_0=\pi/8$ on the grid acts like
a chemical potential $\mu=u_0-\mu_0=-0.416$ in the continuum.
Minnhagen and Wallin predict that the KT line will 
end at $T^*=0.144$ and fugacity $z^*=0.054$, giving a
chemical potential $\mu=T^*\ln z^*=-0.420$. 

To conclude, we show that the 2D neutral CG continues to have an 
ordinary KT transition up to high densities $\rho\sim 0.1$, in 
contrast with recent theoretical estimates.  This result, obtained 
here for the 
square lattice CG, is consistent with recent simulations in the continuum.

One of us (PG) would like to thank Calin Ciordas for interesting 
discussions and the Department of Physics of the University of Rochester 
for a teaching assistantship.
This work was supported by U.S. Department of Energy grant 
DE-FG02-89ER14017.

\end{document}